\documentclass[10pt,conference,letterpaper]{IEEEtran}
\usepackage{times,amsmath,epsfig}
\usepackage[pdfpagelabels,hypertexnames=false,breaklinks=true,bookmarksopen=true,
bookmarksopenlevel=2,bookmarksnumbered=true,pdfpagemode=UseNone,pdfstartview=FitH]
{hyperref}
\usepackage{pifont}
\usepackage[T1]{fontenc}
\title{Significance of Rapid Solutions Development to Business Process
  Management}
\hypersetup{pdftitle={Significance of Rapid Solutions Development to Business
    Process Management},pdfauthor={Steve Kruba},pdfsubject={Rapid BPM}}
\author{
{Steve Kruba}
\vspace{1.6mm}\\
\fontsize{10}{10}\selectfont\itshape
Northrop Grumman\\
3975 Virginia Mallory Drive, Chantilly VA 20151, USA\\
\fontsize{9}{9}\selectfont\ttfamily\upshape
steve.kruba@ngc.com
}
\begin{document}
\maketitle              
\begin{abstract}
  Business process management (\textsc{bpm}) is moving from a niche market into the
  mainstream.  One of the factors leading to this transformation is the emergence of
  very powerful rapid solutions development tools for creating \textsc{bpm} solutions
  (\textsc{bpm~rsd}). It has been widely recognized that this facility is important for
  achieving benefits quickly. Similar benefits are attributed to the agile software
  movement, but \textsc{bpm~rsd} differs in that the objective is to \emph{reduce} the
  need for custom software development. As the \textsc{bpm~rsd} features of some of the
  current business process management suites (\textsc{bpms}) products have matured,
  additional benefits have emerged that fundamentally change the way we approach
  solutions in this space.

\end{abstract}

\vspace{6pt}
\section{Introduction}
Technology, in a traditional sense, is not the differentiator that attracts customers.
In the \textsc{bpm} context, the ``technology'' is about how the extensive functionality
that is required for successfully automating a customer's mission-critical business
processes is applied to solving their problem.

Business users are not information technology (\textsc{it}) experts and have difficulty
relating technical designs to their business needs. Furthermore, most business users
have great difficulty articulating their needs since they have little experience or
involvement working with complex process solutions. This has historically been a major
impediment to creating successful \textsc{bpm} solutions.

Modern \textsc{bpms} products provide a rich application development infrastructure with
significant out-of-the-box capabilities and extensive hooks for customization. This
paper will provide information on these capabilities and the benefits that are
provided. Not only do these capabilities provide a rich environment for building
solutions, but the combination of rapid solutions development and the rich internal
constructs needed to support it amplify a designer's ability to conceptualize these
solutions. By providing the major, base functionality, these products allow architects
and developers to focus on the unique aspects of each solution~--~the issues that make
the difference between successful and unsuccessful projects. Examples will be provided
based on Northrop Grumman's e.POWER\textsuperscript{\textregistered}\footnote{e.POWER is
  a commercial {\sc BPM} product and a registered trademark of the Northrop Grumman
  Corporation.} {\sc bpms} product.

There are over 100 products in the \textsc{bpm} software product market as well as
products servicing other software product segments that have \textsc{bpm} features. A
small subset of these products offer the capabilities described in this paper. The
implications of these capabilities are, perhaps, more significant than have previously
been documented, and affect all aspects of the system development life cycle
(\textsc{sdlc}).  \vspace{-3pt}
\section{BPM RSD Feature Requirements}

\vspace{-1pt} 

\begin{figure}
 \begin{center}
 \vspace{0pt}
  \includegraphics[width=3.4in]{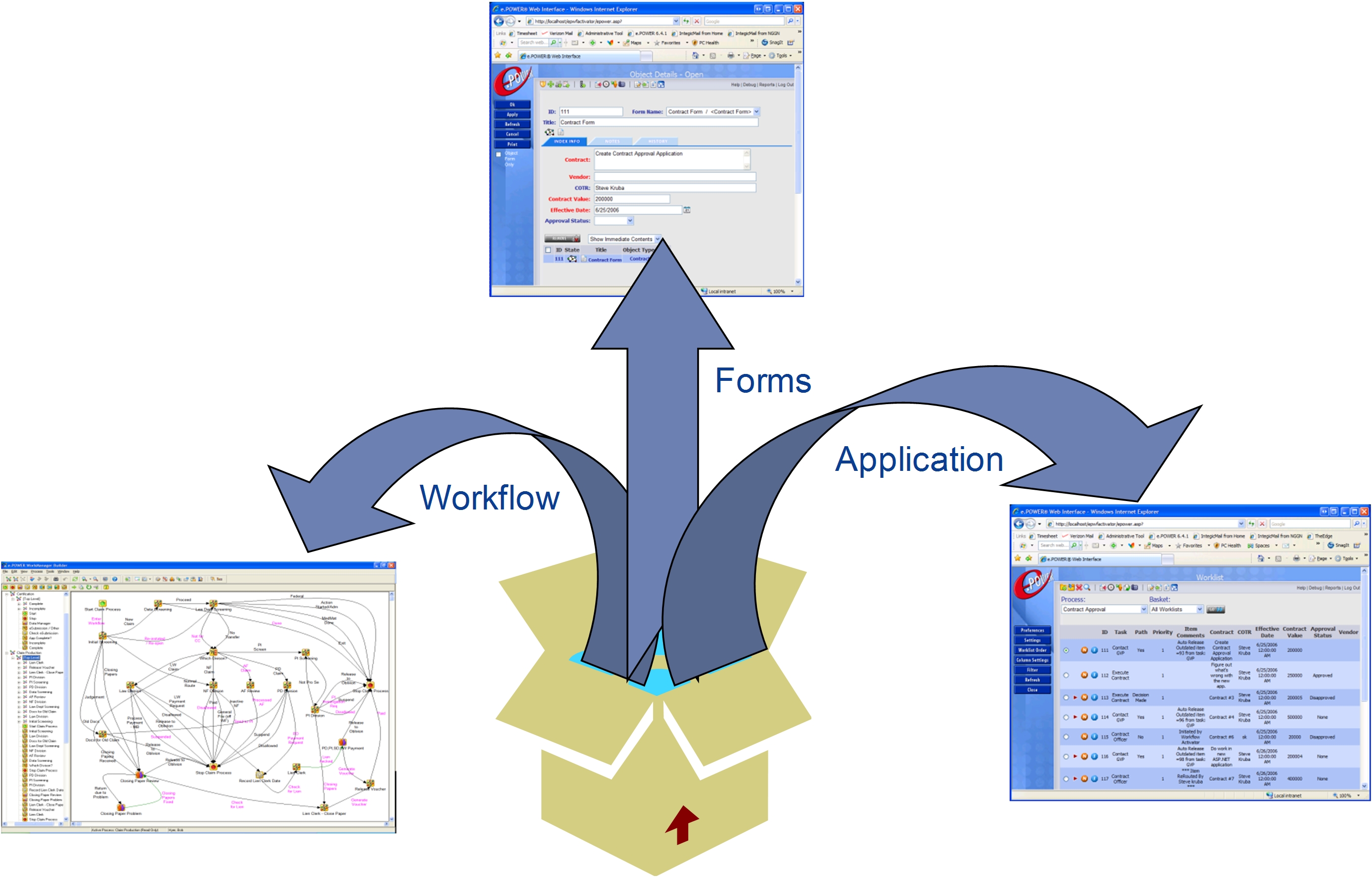}
  \caption{Three Key BPM Components}
 \label{keycomp}
 \vspace{-7pt}
 \end{center}
\end{figure}

\textsc{bpm~rsd} tools focus on providing the three key components required for any
\textsc{bpm} solution: the business process or workflow, an application for doing the
work, and forms as the basis for user interaction. These three components are
illustrated in Figure \ref{keycomp}. The extent to which a particular \textsc{bpm}
product provides these capabilities out-of-the-box is a measure of their
``out-of-the-boxness.'' Keep in mind that not all \textsc{bpm} products have \textsc{bpm
  rsd} toolsets.

\begin{figure*}
 \begin{center}
 \vspace{0pt}
  \includegraphics[width=6.5in]{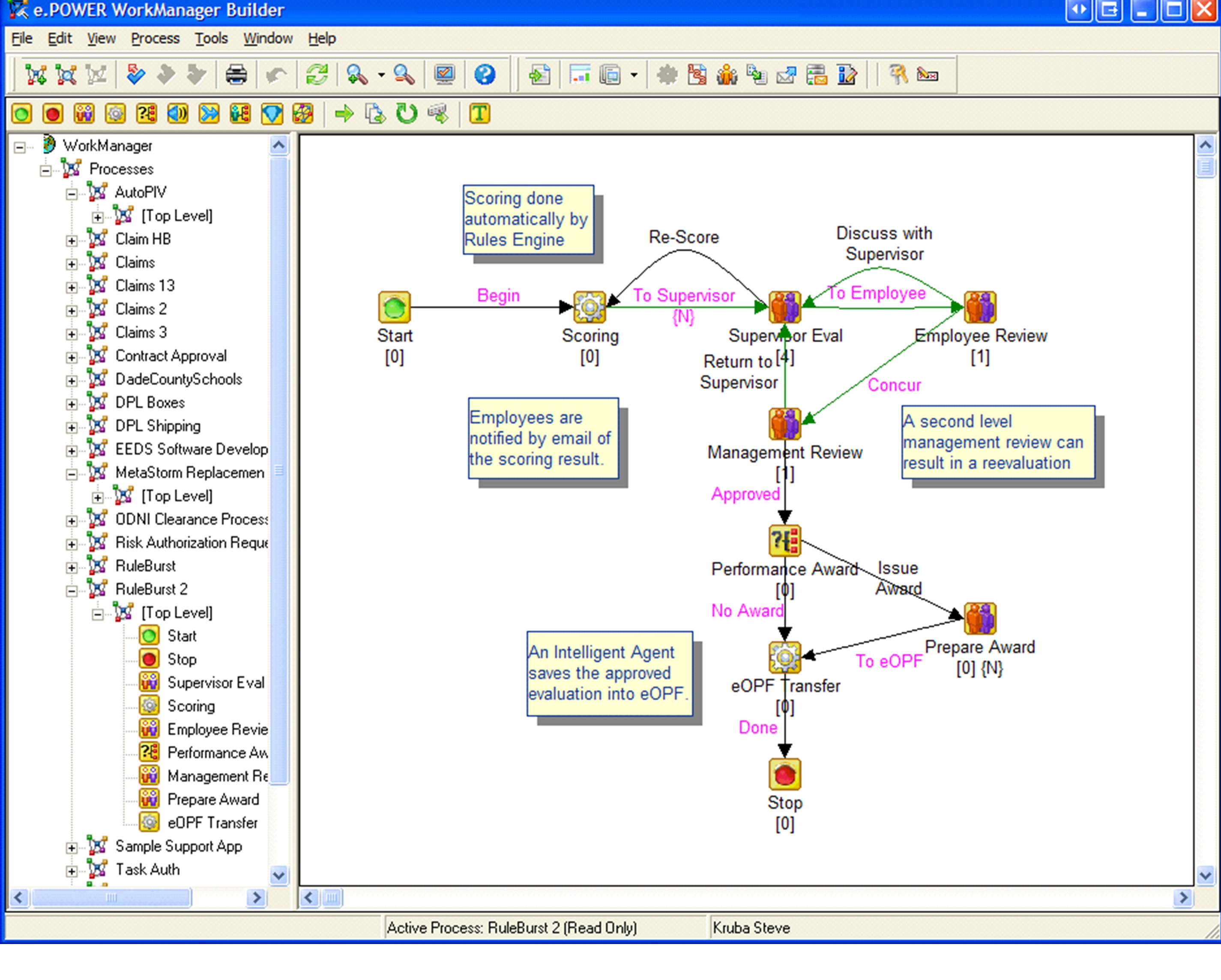}
 \caption{Graphical Workflow Map}
 \label{wfmap}
 \vspace{0pt}
 \end{center}
\end{figure*}

\begin{figure*}
 \begin{center}
 \vspace{0pt}
  \includegraphics[width=6.5in]{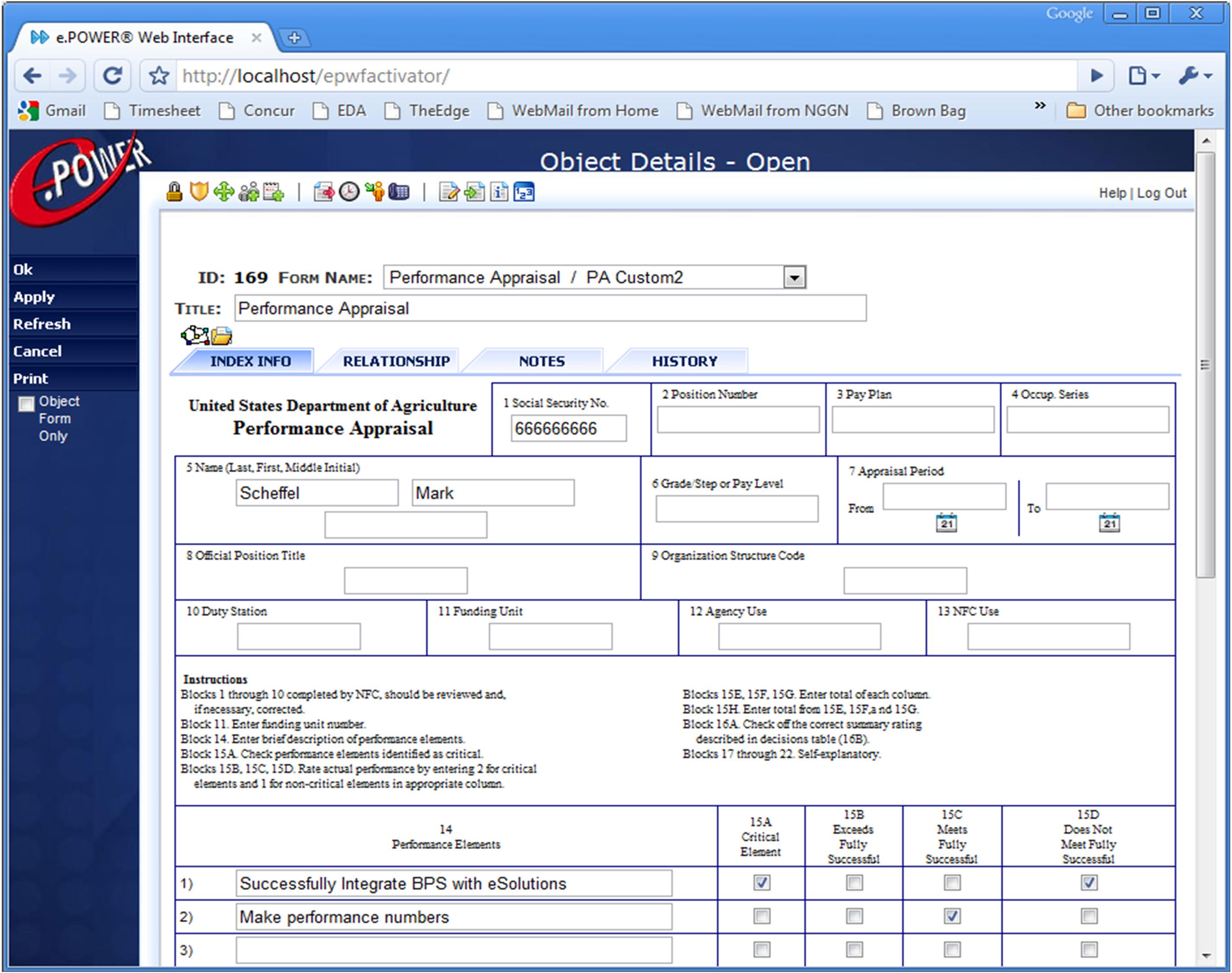}
 \caption{User Interface}
 \label{ui}
 \vspace{-10pt}
 \end{center}
\end{figure*}

Automating a business process involves two key steps:

\begin{dingautolist}{202}

\item Creating an automated representation of the business process~--~see Figure
  \ref{wfmap}. Drag and drop interfaces are the norm with \textsc{bpm~rsd} tools. Note
  that in addition to being a visual representation of the process, it also defines the
  rules for the process in a backend store that is later used by the process engine for
  managing the work. Engines of this type are said to be ``model-driven'' because
  changes to the model directly affect production instances of the process. Other, less
  flexible approaches include configuration-driven and parameterized where a limited set
  of options are baked in by the product vendor. \cite{gartnerthree}

\item Creating an application for users to process their work (see Figure \ref{ui}) that
  is \emph{process-enabled}. For the toolset to be considered a \textsc{bpm~rsd}
  toolset, the user interface should be a byproduct of the application definition
  process~--~a declarative process rather than a programming exercise.  While it is
  important to automatically generate the user interface, it is also important to
  provide customization hooks needed to tweek the interface, since rarely is the
  one-size-fits-all approach adequate.

\end{dingautolist}

\section{Agility}

The primary purpose of process automation is process improvement. Complex business
processes are constantly changing~--~with or without explicit direction. Factors such as
changing business environments, government regulation, and competition are major drivers
of these changes, necessitating changes in the support systems.

The traditional life-cycle development approaches to custom development are seriously
challenged to support these dynamics.\footnote{Cantara, p.7.} Historically, requirements
documentation, detailed systems design, development, and implementation could easily
take 12 to 18 months to deliver a complex solution, during which time the business
requirements may have changed significantly enough to require additional iterations
prior to implementation.

\emph{Agility} has become a popular term for describing the flexibility needed by
organizations to operate in today's dynamic environments. Agility is a natural
by-product of \textsc{bpm~rsd} toolsets. Agility is a critical feature of \textsc{bpms}
products.

Agility in the \textsc{bpm} context is similar to, but not the same as the agile
software development methodology, typically used in an iterative process for creating
custom software. Agility in the \textsc{bpms} space is more about using the built-in
capabilities of the \textsc{bpm} product to \emph{avoid} having to write custom
software. Custom software is needed as part of the creation process for most
\textsc{bpm} solutions, but whenever it can be avoided, the resulting solution is less
expensive and has fewer defects and lower risk. In order to differentiate this process
from rapid applications development (\textsc{rad}) approaches, I have coined the term
``\textsc{bpm~rsd}.''

Another related software engineering concept is model-driven development. These
techniques often include a framework in which software is developed, providing a
powerful facility for leveraging the assets within the framework for reuse.

The concept of `models' is critical to \textsc{bpm} products where the software
architecture creates models of the organization's business operations~--~a key example
being the model of the operational aspects of the business being encapsulated in the
graphical process map. But as in the agile space, the model-driven development space is
critically different from \textsc{bpm~rsd} in one respect: it is meant to either
generate source code or provide a framework within which source code is
written. \textsc{bpm~rsd} models are run-time models as well as design-time models and
are designed to reduce the need to write custom software.

So how are \textsc{bpm~rsd} tools different? For agile or model-driven development,
source code must be recompiled and redeployed when changes are made. Although this might
be done automatically, it does not produce a clean transition in production
environments; i.e., it is not seemless to an operating business
process. \textsc{bpm~rsd} products, however, store the semantics of process definitions
in a repository~--~often a relational database~--~and execution engines dynamically
drive production instances from this repository. Changes to the repository using the
\textsc{bpm~rsd} tools directly effect operational changes.

Another key distinction between frameworks and \textsc{bpm~rsd} tools is that frameworks
require skilled software developers to ``wire up'' the framework in order to achieve the
benefits. Frameworks are analogous to e.POWER's \textsc{api}'s plus our solution
paradigm, but in e.POWER our framework has been pre-wired so that many of the technical
requirements have already been resolved, allowing less skilled staff, or in some cases
such as our process designer, non-technical staff, to contribute to solution
development. Frameworks also require that individual wires be ``soldered'' into the
solution. \textsc{bpm~rsd} tools eliminate the need to do so and eliminate the
possibility of \emph{neglecting} to do so, insuring that critical functionality such as
auditability, searchability, etc., mentioned in the Object Types section of this paper,
is included automatically.

In a very real sense, \textsc{bpm~rsd} tools are pre-wired, or pre-compiled frameworks.

\section{Out-of-the-Boxness}

Similar approaches have arisen over the years in other business software
categories. Typically packaged as products to offset the increased cost of producing
these solution sets, these products consist of design tools that are largely
configuration-driven and produce robust implementations. Such solution sets exist in the
enterprise resource planning (\textsc{erp}) space, customer resource management space
({\sc crm}), as well as the \textsc{bpm} space.  Each of the design-time toolsets has
unique characteristics. One of the key differentiators is how much functionality is
delivered ``out-of-the-box'' and how much requires custom software development.

This ``out-of-the-boxness'' has significant benefits beyond the obvious advantage of
creating solutions quickly. Successful \textsc{bpm} solutions must be customizable to
each organization's unique requirements. Gathering those requirements through
traditional documentation approaches can be cumbersome and slow and produces paper-based
models to validate the requirements~--~an imperfect model at best.

\textsc{bpm~rsd} tools provide working models of the solution in days rather than weeks
or months. The significant user interfaces and workflow needed for requirements
validation can be mocked up very quickly, providing a significant portion of the
solution in a totally objective fashion~--~via working software. Select portions of the
solution such as legacy systems integrations might be delayed to a later phase in order
to minimize the impact on requirements gathering or to accelerate implementation in
order to achieve operating efficiency gains earlier in the process. Iterating with
prototypes help business people and \textsc{it} staff objectify the end-product more
quickly and accurately, greatly increasing the likelihood of a successful
implementation.

The obvious advantage of \textsc{bpm~rsd} development is also important: rapid
implementations. Lengthy requirements efforts of many months suffer from the modern-day
problem of a rapidly changing business context. How often have we seen a system that was
well-designed and executed, but outdated by the time it was deployed?  \textsc{bpm~rsd}
approaches reduce that risk.

It is important to note that these prototypes are not throw-aways. To the extent they
accurately reflect the underlying requirements, they become part of the final production
solution. The key is that the tools used to develop proofs-of-concept, prototypes, and
production systems are \emph{the same tools.}

\section{Object Types}

Automation of business systems require creation of software modeling constructs that
represent the key business objects in the problem domain. We refer to these constructs
as \emph{object types}. These object types map to the real-world business objects in the
same way that classes relate to class instances in object-oriented programming
languages. Object types could be thought of as index fields -- or metadata on steroids.

Effective \textsc{bpm~rsd} tools require a rich structure for creating process-enabled
applications of any complexity. This generic structure, while necessary to support the
user interfaces generated, is also very effective at helping analysts conceptualize the
ultimate solution.

In tools such as the e.POWER Activator designer, object types define the characteristics
of the real-world business components that make up the solution and object instances
model the actual instances of those business objects. A by-product of this approach is
that objects and object types inherit many useful properties from the e.POWER Activator
infrastructure, features that might not be provided if the solution were created in a
custom development effort. All e.POWER Activator objects are inherently editable and
searchable and all activity against them is auditable. The rules applied at design time
help to insure data integrity.

For an equal employment opportunity (\textsc{eeo}) application, object types might
include complainant, class action lawsuit, and investigator. For a training solution,
object type definitions would be needed for student, training class, and possibly
training facility if the application was designed to model any of the facility
behaviors. This direct mapping between \textsc{it} constructs and business constructs
greatly facilitate communication between \textsc{it} staff and business staff and
simplifies solution conceptualization.

\section{BPM RSD Benefits}

Summarizing what we've discussed so far, \textsc{bpm~rsd} tools provide the following
key benefits over more traditional approaches.

\begin{dingautolist}{202}

\item \emph{Requirements gathering}. Providing a prototype solution early in the
  requirements gathering process helps end users understand the possibilities and helps
  to shape their expectations.

\item \emph{Requirements validation}. Working prototypes allow end users to understand
  exactly what they are getting. It is very difficult for end users to visualize how the
  software will affect how they work from paper documentation. Prototypes help to
  eliminate the perennial problem of ``that's what I asked for but not what I need.''

\item \emph{Analysis and design}. Prototypes also assist designers in visualizing what
  the ultimate solution can and should look like. \textsc{bpm~rsd} capabilities allow
  them to draw from a toolbox of components that include ``nice-to-have'' features that
  might otherwise be omitted from the solution.

\item \emph{Documentation}. All solutions, whether using \textsc{bpm~rsd} or more
  traditional approaches, require many forms of documentation: documentation for project
  approval, documentation for design reviews, documentation for the quality assurance
  process, etc. Having working prototypes early in the process makes all forms of
  documentation significantly easier to produce and much more effective. The clarity
  provided by actual screen-shots, process maps, and relational designs (necessarily
  generated automatically by \textsc{bpm~rsd} tools) benefits all participants in the
  review process, from end-users to the approving management staff.

\item\emph{Implementation}. Implementation is the obvious area where \textsc{bpm~rsd} is
  valuable, allowing customers to achieve the benefits more quickly and less expensively
  than through traditional approaches.

\item \emph{Quality assurance}. The quality of a solution constructed through pre-built
  components is clearly higher since the out-of-the-box features have been refined
  through pre-existing, broad-based customer usage. New customers benefit from defects
  that were identified and resolved by other customers. Additionally, for each product
  release, the software is quality-checked through an independent process. This allows
  \emph{project} quality teams to focus on the customizations~--~the area most likely to
  introduce software defects.

\item \emph{Maintenance}. The area of maintenance aligns with the notion of
  agility~--~being able to modify the production solution to adapt to changing
  conditions. The tools that facilitate rapid creation are typically the same tools used
  to update the solution as needs change over time.

\item\emph{Risk}. The \textsc{bpm~rsd} approach reduces risk in virtually all phases of
  the development effort. The functioning prototypes reduce the risk of building a
  \emph{good} solution that is the \emph{wrong} solution. Analysis and design are
  improved through the objectivity of these same functioning prototypes, reducing the
  risk of an incorrect design. Quality is improved as noted above and therefore reduces
  the risk of poor quality. Implementation, operation and maintenance are likewise
  facilitated, reducing risk in their respective areas as well.

\end{dingautolist}

\textsc{bpm~rsd} products affect all aspects of the system development life cycle and,
as we shall describe in the next section, fundamentally change the way we approach
solutions.

\section{A BPM RSD Methodology}

As stated earlier, these new capabilities suggest a new approach to solution
creation. \cite{gartnerchange} Rather than the traditional waterfall approach of
requirements, design, development, and implementation, our approach is to use the
following roadmap when engaging new customers. This approach is iterative: very similar
to an agile software development approach, but the final result is achieved largely
through model-manipulation rather than programming.

\begin{dingautolist}{202}

\item Request existing documentation from the business users very early in the
  requirements gathering process.
	
	\begin{enumerate}
   \item A Visio diagram or a description of the business process is the starting point
     for creating the process map.
   \item Copies of key forms provide templates for some of the user-interfaces as well
     as the data fields needed for the application.
   \end{enumerate}

  \item Prototype the solution using the \textsc{bpm~rsd} tools.

	\begin{enumerate}
   \item Use the graphical process designer to draw the business process which is more
     than visual: it encapsulates the business rules that drive the process. This
     graphical representation is critical to making sure \textsc{it} and the business
     users agree on the process details.
   \item Use a declarative application builder for application creation.
   \item Use a security manager for defining security profiles, often integrated with an
     existing \textsc{ldap} repository.
	\end{enumerate}

 \item Present the solution to the business users to refine the requirements and the
   solution.
   
   \begin{enumerate}
   \item Iterate on changes to the process diagram in interactive design sessions.
   \item Modify the application in interactive design sessions.
   \end{enumerate}

 \item Put the solution into production, often in phases to accelerate the initial
   benefits.
   
   \begin{enumerate}
   \item Create documentation from the prototype to support the organization's
     vetting process.
   \item Get something into production quickly to get immediate benefits.
   \item Defer complex integrations until later phases if possible.
   \end{enumerate}

\end{dingautolist}

This relatively simple formula is significantly more effective than alternative
methods. Business users are able to react to high-fidelity prototypes rather than paper
representations. \textsc{bpm~rsd} tools make it feasible to rapidly evolve the
solutions~--~interactively in design sessions with the end-users.

\section{Conclusions}

\textsc{bpm} rapid solutions development tools make it possible to construct business
process solutions much more quickly and effectively than in the past. End-users are able
to interact with designers in an expressive environment that allows them to model the
solution with the actual tools used to create the solution. Rapid prototyping is a key
to this approach and allows developers to build the solution that the business users
need to satisfy their \emph{actual} requirements~--~rather than the ones they ``asked
for.''

This represents a major step forward in producing effective solutions. This approach
results in lower risk of producing an ineffective solution and reduces defect rates by
minimizing the amount of custom coding required to produce the solution. The end result
is a much higher probability of successful projects. The \textsc{bpms} software market
serviced by model-driven \textsc{bpm~rsd} tools may be unique in the \textsc{it}
software industry.

\vspace{23pt}
\noindent{\small\emph{\textbf{Steve Kruba} is Chief Technologist for e.POWER
  product development and a Northrop Grumman Technical Fellow.}}

\end{document}